\documentclass[11pt]{article}
\usepackage{moriond,epsfig}

\def\mco{\multicolumn}

\def\be{\begin{equation}}
\def\ee{\end{equation}}
\def\bea{\begin{eqnarray}}
\def\eea{\end{eqnarray}}

\newcommand{\mbc}{m_{\rm bc}}
\newcommand{\eb}{E_{\rm beam}}
\newcommand{\de}{\Delta E}
\newcommand{\bb}{B\bar{B}}
\newcommand*{\brkspp}{53.2\pm11.3\pm9.7}
\newcommand*{\brkskk}{34.8\pm6.7\pm6.5}
\newcommand*{\brkskp}{9.3}

\newcommand*{\brksppkstpi}{13.5^{+5.0}_{-4.4}\pm{2.9}}
\newcommand*{\brksppkstxpi}{22.9^{+8.7}_{-8.0}\pm{6.0}}

\newcommand*{\brkskkphiks}{6.4^{+3.0}_{-2.6}\pm1.3}
\newcommand*{\brkskkfxks}{20.4^{+5.3}_{-4.9}\pm3.8}

\begin{document}
\vspace*{4cm}
\title{RARE $B$ DECAYS AT BELLE}

\author{T.J. GERSHON \\ (for the Belle collaboration)}

\address{KEK, High Energy Physics Research Organization, 1-1 Oho, Tsukuba, Ibaraki, 305-0801 Japan}

\maketitle\abstracts{
Results on a number of rare hadronic $B$ decays are presented.
A data sample of $29.1\ {\rm fb}^{-1}$ accumulated using the Belle detector at the KEKB asymmetric $e^+e^-$ collider operating at the $\Upsilon(4S)$ resonance is used. 
All results are preliminary unless indicated otherwise.
}

\section{Introduction}
Belle is an experiment designed to make measurements of time-dependent $CP$ asymmetries in $B$ physics.  
Important measurements of these phenomena have recently been reported.\cite{higsag}  
In addition, owing to the unparalleled performance of KEKB,\cite{kekb}
and the excellent performance of the Belle detector, 
numerous interesting rare $B$ decay modes can be studied.

The Belle detector is described in detail elsewhere.\cite{bellenim}  
It consists of 
a silicon vertex detector, 
a central drift chamber (CDC), 
aerogel Cerenkov counters (ACC), 
time-of-flight scintillation counters (TOF) 
and an electromagnetic calorimeter made of CsI(Tl) crystals 
enclosed in alternating layers of resistive plate chambers and iron for $K_L/\mu$ detection and to return the flux of the 1.5 T magnetic field.  
The responses of the CDC, ACC and TOF are combined to provide clean identification of charged kaons, pions and protons, crucial for rare hadronic $B$ decays.
Another important feature of rare decay analyses, is the suppression of background from $e^+e^- \to q\bar{q}$ (continuum) events.  
One commonly used tool is a likelihood ratio
combining information from the reconstructed B direction and a Fisher
discriminant of modified Fox-Wolfram moments.\cite{sfw}

Signal candidates are identified using two kinematic variables:
 $\mbc=\sqrt{\eb^2 - (\Sigma {\mathbf P}_{i})^2}$ and 
 $\de = \Sigma E_{i} - \eb$.
Signal yields are obtained by fitting the $\de$ distributions, since this method allows separation of the continuum and $\bb$ backgrounds from the signal.

\section{Three Body Charmless Hadronic Decays}
Belle has recently published results on three body charmless hadronic decays $B^+ \to K^+ h^+ h^-$,\cite{cc} where $h=K,\pi$.\cite{khh}
These results include the first observation of a $B$ decay to a scalar pseudoscalar final state, $B^+ \to f_0(980)K^+$.  
It is of interest to look for similar phenomena in the neutral channel, to further investigate the $b \to s$ penguin transitions which mediate these decays.
In addition, these modes may in future be used to measure direct CP violation.

\begin{figure}
\resizebox{0.24\textwidth}{!}{\includegraphics{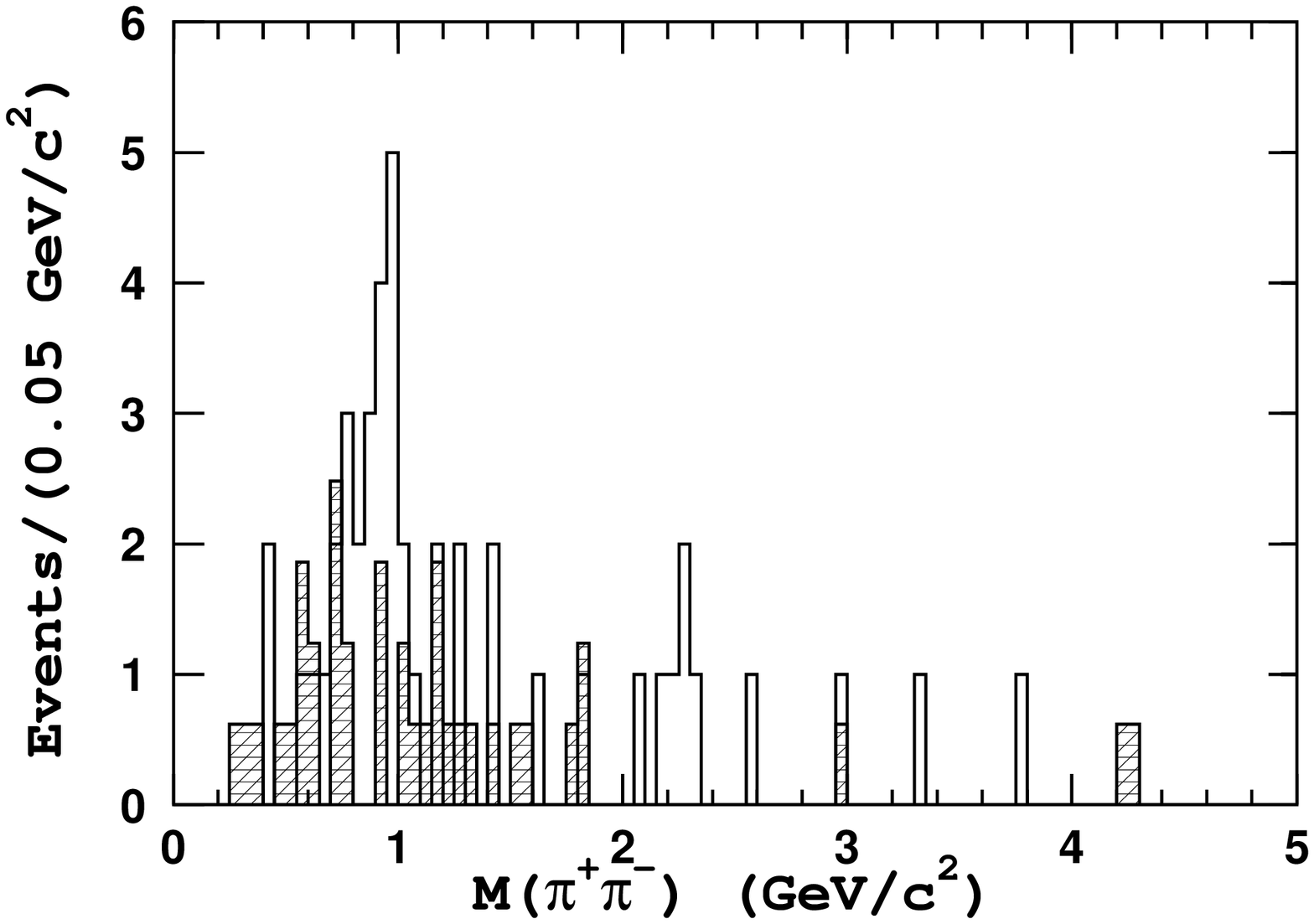}}
\hfil
\resizebox{0.24\textwidth}{!}{\includegraphics{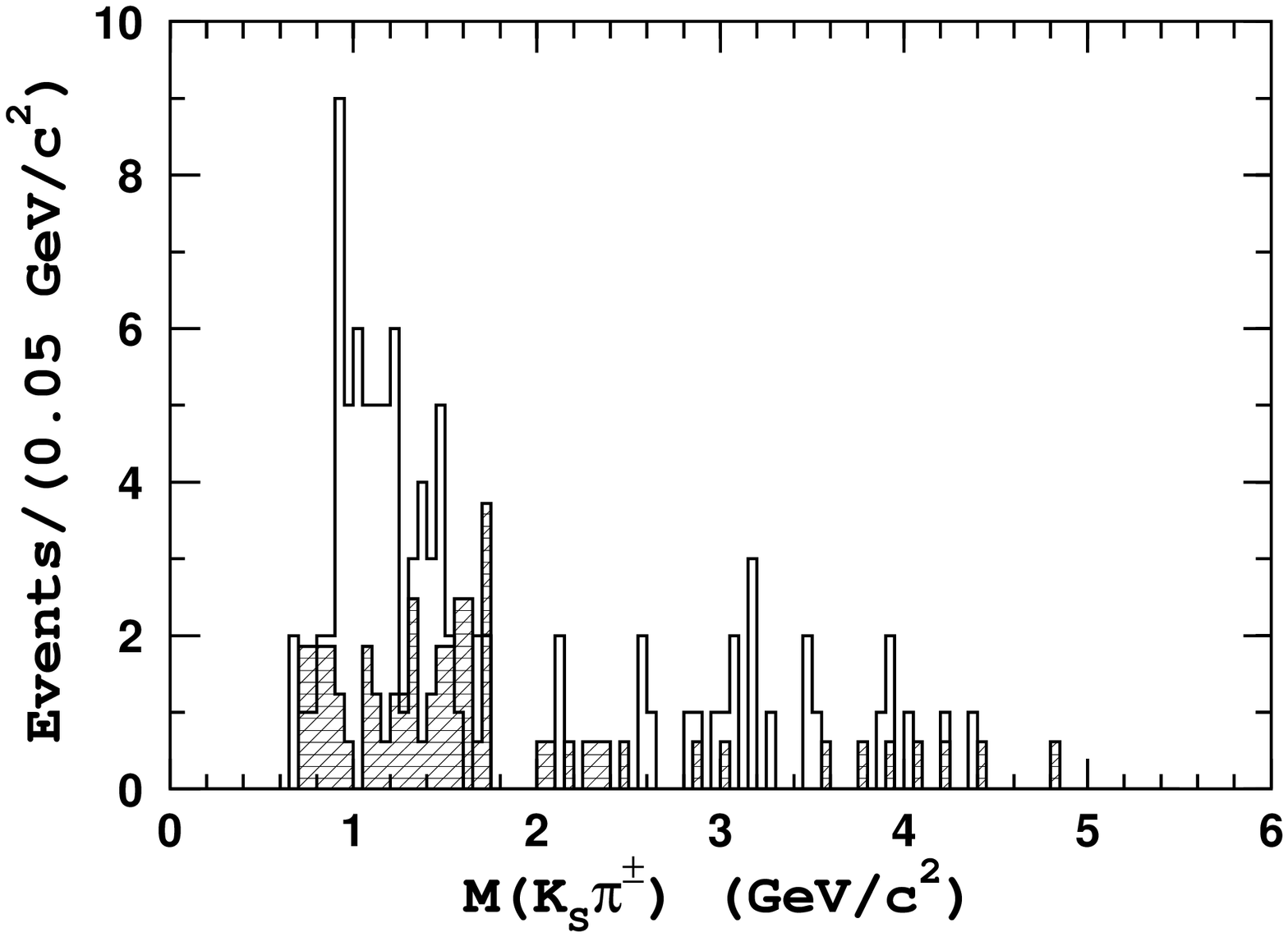}}
\hfil
\resizebox{0.24\textwidth}{!}{\includegraphics{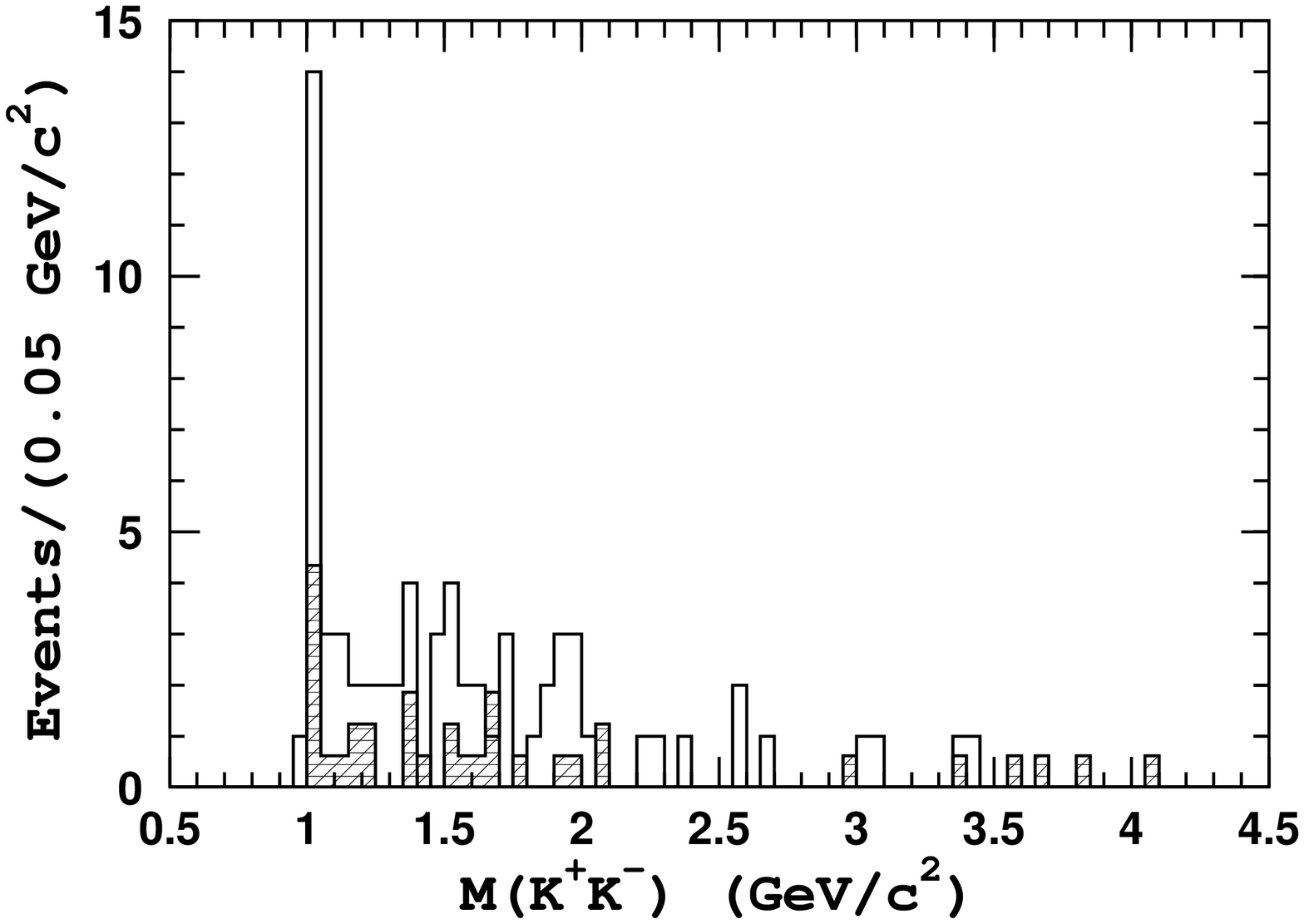}}
\hfil
\resizebox{0.24\textwidth}{!}{\includegraphics{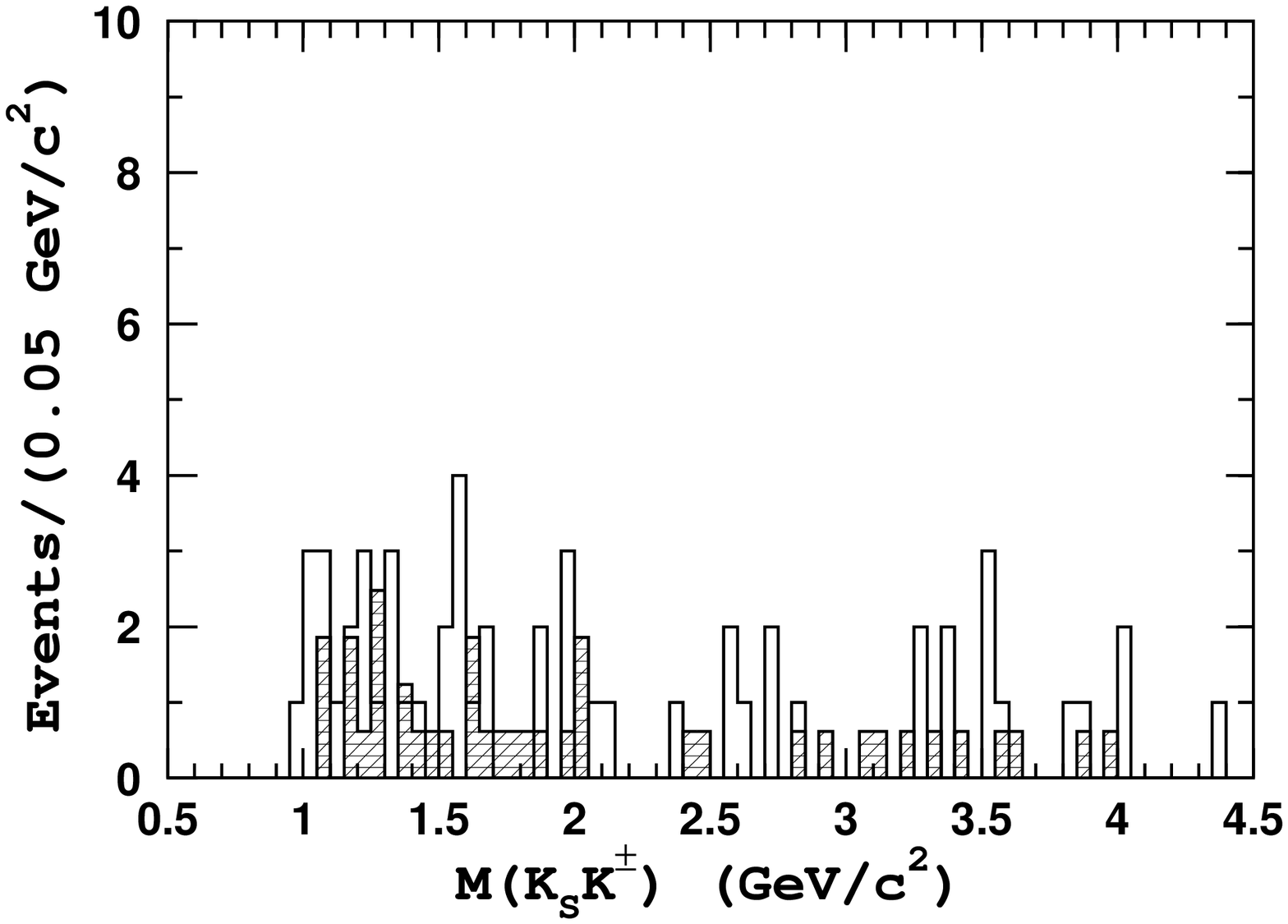}}
\caption{
  From left to right: 
  $\pi^+\pi^-$ invariant mass for $B \to K^0\pi^+\pi^-$ candidates with $m_{K_S\pi^{\pm}} > 2.0~{\rm GeV}/c^2$; 
  $K_S\pi^{\pm}$ invariant mass for $B \to K^0\pi^+\pi^-$ candidates with $m_{\pi^+\pi^-} > 1.5~{\rm GeV}/c^2$; 
  $K^+K^-$ invariant mass for $B \to K^0K^+K^-$ candidates; 
  $K_S K^{\pm}$ invariant mass for $B \to K^0K^+K^-$ candidates with $m_{K^+K^-} > 2.0~{\rm GeV}/c^2$. 
  In each case the shaded histogram shows the background measured in a $\Delta E$ sideband.
  \label{fig:khh}
}
\end{figure}

The results are summarized in table~\ref{tab:khh}.  
Branching fractions (BFs) for $B^0 \to K^0 \pi^+ \pi^-$ and $B^0 \to K^0 K^+ K^-$ have, without any assumptions about intermediate mechanisms, been measured for the first time.  
BFs for intermediate resonances are extracted in each mode using a simultaneous fit to the projections of the Dalitz plot, shown in figure~\ref{fig:khh}.
The results are consistent with those from charged $B$ decays, albeit with larger statistical error (due to ${\cal B}(K^0 \to \pi^+\pi^-)$).  
The vector pseudoscalar decay $B^0 \to K^{*-}(892)\pi^+$ is observed.

\begin{table}[t]
\caption{
  BFs and 90\% confidence level (CL) upper limits, 
  in units of $10^{-6}$, for $B \to K_S h^+ h^-$ and intermediate resonances.
  \label{tab:khh}
}
\vspace{0.2cm}
\begin{center}
\begin{tabular}{|l|cc|cc|c|} \hline
Mode                       & \mco{2}{c|}{$K^0\pi^+\pi^-$}       & \mco{2}{c|}{$K^0 K^+  K^-$}    & $K^0 K^{\pm} \pi^{\mp}$ \\
${\cal B} \to K^0 h^+ h^-$ & \mco{2}{c|}{$\brkspp$}             & \mco{2}{c|}{$\brkskk$}         & $< \brkskp$ \\
Resonance                  & $K^*(892)^-\pi^+$ & $K_X(1400)^-\pi^+$ & $\phi(1020) K^0$ & $f_X(1500) K^0$ & \\ 
${\cal B} \to R_{\to hh}h$ & $\brksppkstpi$  & $\brksppkstxpi$  & $\brkskkphiks$ & $\brkskkfxks$ & \\
\hline
\end{tabular}
\end{center}
\end{table}

\section{Two Body Charmless Hadronic Decays}
Two body charmless decays provide a rich variety of $B$ physics.\cite{hhtheory}
CP violation may be observed in the time-dependence of $B \to \pi^+\pi^-$ decays,
whilst due to penguin pollution, isospin analysis of the $\pi\pi$ system 
will be required to extract the Unitary Triangle angle $\phi_2$.
The relative sizes of other $B \to hh$ modes provide information 
about the magnitudes of tree, penguin and other diagrams which contribute.
Also, these modes may be used to search for direct CP violation, and to extract $\phi_3$.
Belle has previously published results on these modes using $10.4\ {\rm fb}^{-1}$\cite{hh};
 updated results are shown in table~\ref{tab:hh}. 
There is evidence of $B^+ \to \pi^+\pi^0$ with a significance of $3.5\sigma$, 
and a $2.2\sigma$ hint for $B \to \pi^0\pi^0$.  
A possible CP asymmetry in $B^+ \to K^0 \pi^+$ is found.  
Such an effect would be a clear indication of new physics if confirmed.\cite{neubert}
The ratio $\Gamma(\pi^+\pi^-)/2\Gamma(\pi^+\pi^0)$ is found to be 
$0.40 \pm 0.15 \pm 0.05$,
indicating large tree-penguin interference.

\begin{table}[t]
\caption{
  BFs and 90\% CL upper limits, in units of $10^{-6}$, for $B \to h h$ decays.  
  CP asymmetries are given for flavour specific final states.
  \label{tab:hh}
}
\vspace{0.2cm}
\begin{center}
\begin{tabular*}{\textwidth}{|l|}
\hline
\begin{tabular*}{0.973\textwidth}{@{\hspace{-0.5mm}}l@{\extracolsep{\fill}}c@{\extracolsep{\fill}}c@{\extracolsep{\fill}}c@{\extracolsep{\fill}}c} 
Mode         & $K^+  \pi^-$ & $K^+  \pi^0$ & $K^0  \pi^+$ & $K^0  \pi^0$ \\

${\cal B}(B \to hh)$    & $~2.18 \pm 0.18 \pm 0.15$ & $~1.25 \pm 0.24 \pm 0.12$ & $~1.88 \pm 0.30 \pm 0.15$ & $~0.77 \pm 0.32 \pm 0.16$ \\
${\cal A}_{CP}$         & $-0.06 \pm 0.08 \pm 0.01$ & $-0.04 \pm 0.19 \pm 0.03$ & $~0.46 \pm 0.15 \pm 0.02$ & \\
\end{tabular*}
\\
\hline
\begin{tabular*}{0.973\textwidth}{@{\hspace{-0.5mm}}lcccccc}
Mode         &  $\pi^+\pi^-$ & $\pi^+\pi^0$ & $\pi^0\pi^0$ & $K^+  K^-$   & $K^+  K^0$   & $K^0  K^0$ \\
${\cal B}(B \to hh)$  & $0.51 \pm 0.11 \pm 0.04$ & $0.70 \pm 0.22 \pm 0.08$ & $< 0.56$ & $< 0.05$ & $< 0.38$ & $< 1.3$ \\
${\cal A}_{CP}$ &  & $0.31 \pm 0.31 \pm 0.05$ & & & & \\
\end{tabular*}
\\
\hline
\end{tabular*}

%
%
\end{center}
\end{table}

\section{$\phi_3$ Program}
Whilst $\phi_3$ may be extracted from measurements of $A_{CP}$ in $B \to hh$ decays,
the contributing diagrams introduce significant theoretical uncertainty in the extraction.
A cleaner method to obtain $\phi_3$ is from the CP asymmetry in $B^+ \to D_{CP}K^+$ decays.\cite{dcpk}
The measured CP asymmetries when the $D$ meson is reconstructed in a CP=$+1$ eigenstate,  a CP=$-1$ eigenstate and a CP non-eigenstate are $0.29 \pm 0.09 \pm 0.04$, $-0.22 \pm 0.24 \pm 0.04$ and $0.00 \pm 0.09 \pm 0.04$ respectively.
The current statistics are too few to draw any conclusions regarding $\phi_3$.

\section{Charmless Baryonic Decays}
Belle has recently published the first observation of a $b \to s$ penguin transition with baryons in the final state, $B^+ \to p\bar{p}K^+$.\cite{ppk}
In this mode, an excess of events (compared to a phase space prediction) is seen at low invariant mass of the $p\bar{p}$ pair.  
Whilst models which can explain such an excess exist,\cite{ppk_theory} 
further study with more data is needed to understand the mechanisms at work.
Searches for two body charmless baryonic final states 
have so far yielded only upper limits.\cite{pp}

\section{Charmed Baryonic Decays}
To investigate the mechanism of baryonic decays of $B$ mesons,\cite{charmb} it is instructive to study charmed decays, where the BFs are larger.  
The modes $\bar{B}^0 \to \Lambda_c^+\bar{p}\pi^+\pi^-$ \& $B^- \to \Lambda_c^+\bar{p}\pi^-$ 
have been observed by CLEO.\cite{charmbcleo} 
From the same final states, the resonant decays $\bar{B}^0 \to \Sigma_c^{++/0}(2455/2520)\bar{p}\pi^{-/+}$ are observed.\footnote{The significance of the $\Sigma_c^{0}(2520)\bar{p}\pi^{-/+}$ signal is $2.4\sigma$; the other modes have significance greater than $5\sigma$.}
Evidence for $B^- \to \Sigma_c^0(2455)\bar{p}$ is found with $4.7\sigma$ significance.

\section{Colour Suppressed Decays}
\begin{figure}
\resizebox{0.28\textwidth}{!}{\includegraphics{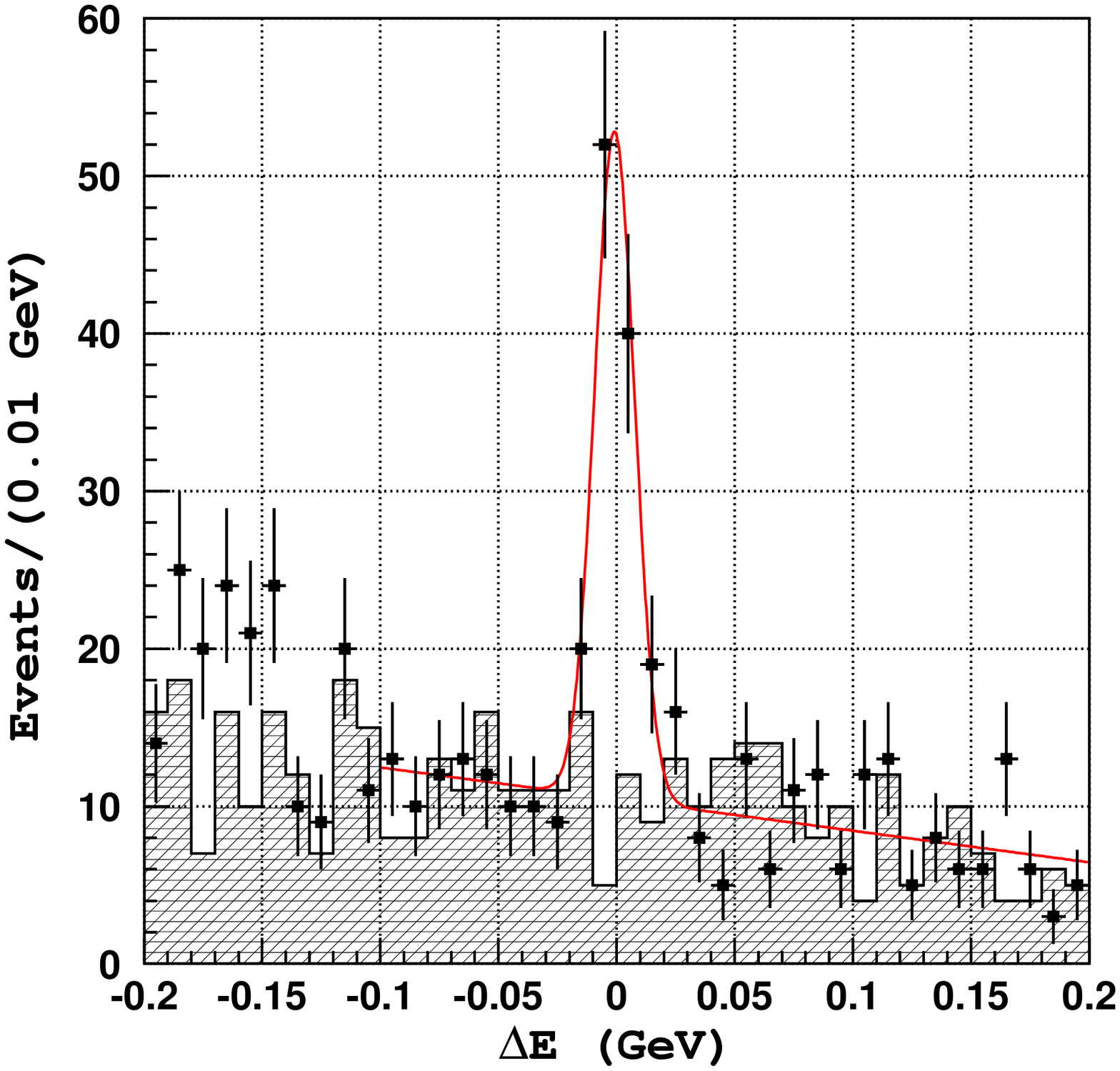}}
\hfil
\resizebox{0.28\textwidth}{!}{\includegraphics{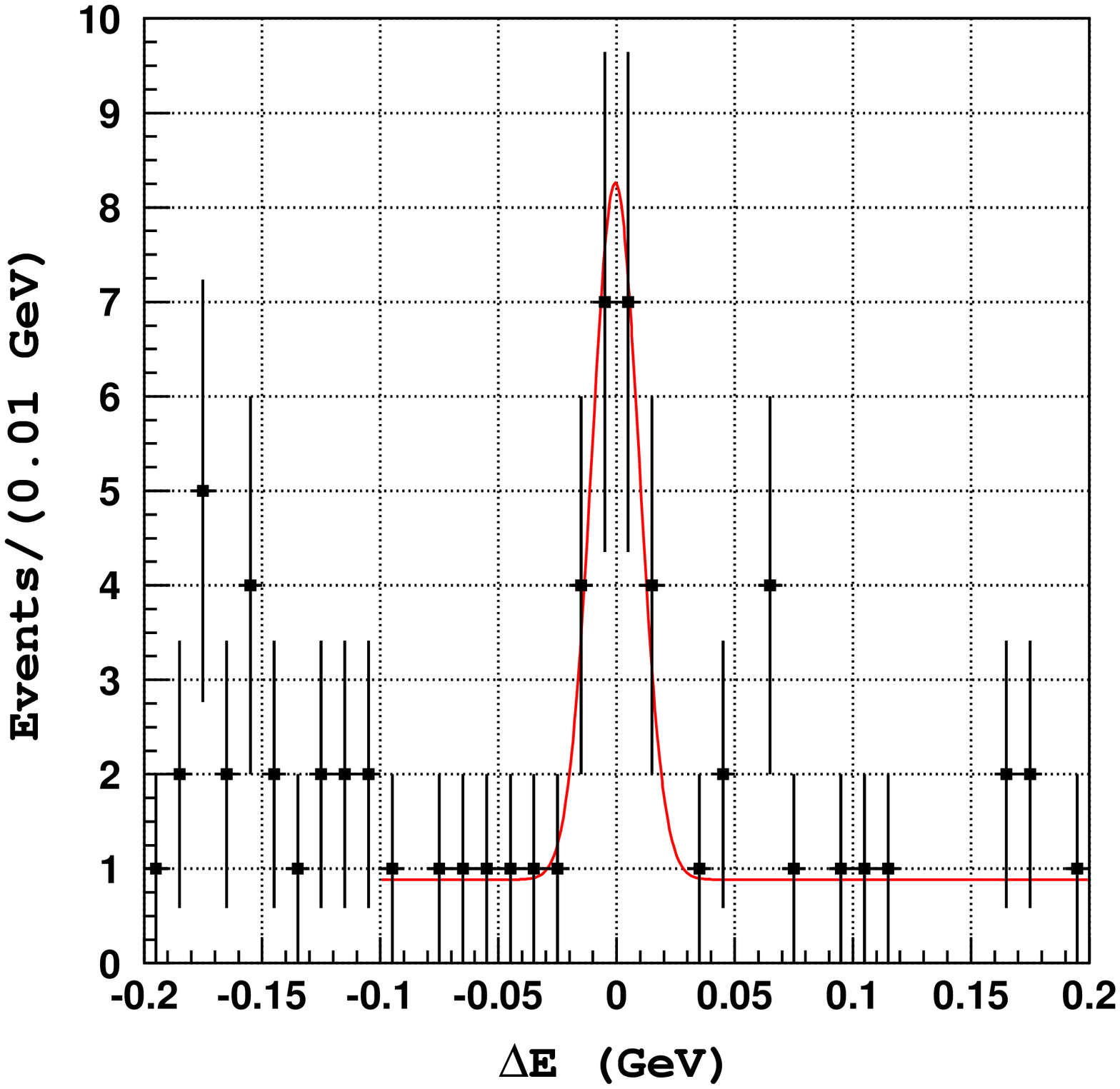}}
\hfil
\resizebox{0.28\textwidth}{!}{\includegraphics{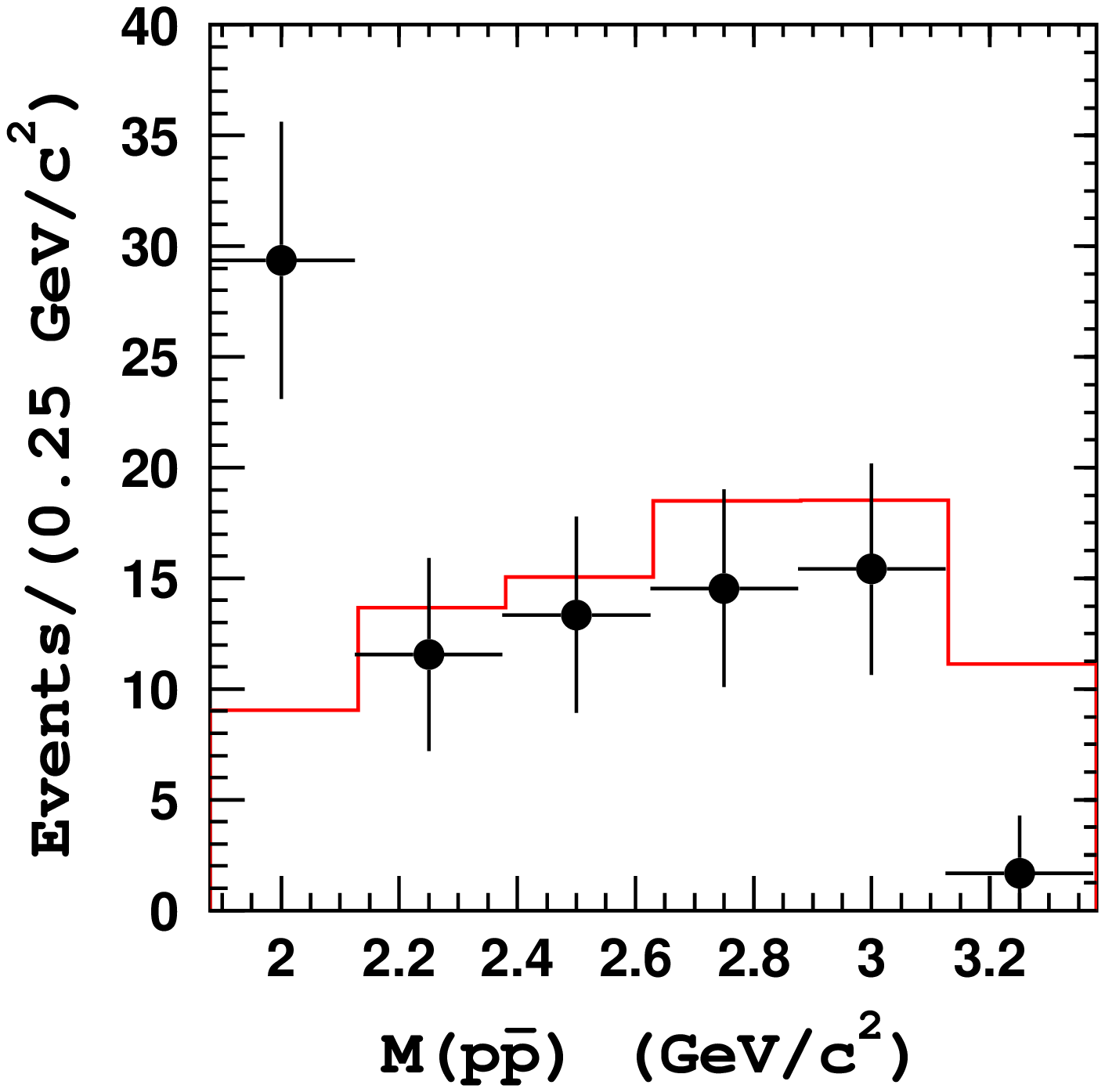}}
\caption{
  From left to right:
  $\Delta E$ distributions for $B \to D^{0}p\bar{p}$ 
  and for $B^0 \to D^{*0}p\bar{p}$ candidates, 
  and the $p\bar{p}$ invariant mass distribution for $B \to D^{0}p\bar{p}$ candidates.
  \label{fig:dpp}
}
\end{figure}

Colour suppressed $B \to D^{*(0)}h^0$ decays have recently been observed.\cite{d0h0}
Figure~\ref{fig:dpp} shows signal peaks for the colour suppressed modes $B^0 \to D^{(*)0}p\bar{p}$.  
This is the first observation of these modes; 
the BFs are 
${\cal B}(B^0 \to D^{0}p\bar{p})  = (1.18 \pm 0.15 \pm 0.16) \times 10^{-4}$ and 
${\cal B}(B^0 \to D^{*0}p\bar{p}) = (1.20^{+0.33}_{-0.29} \pm 0.21) \times 10^{-4}$.
The invariant mass of the $p\bar{p}$ pair for $B \to D^{0}p\bar{p}$ candidates is also shown in figure~\ref{fig:dpp}; a peak at low mass is observed, as was the case in the $B^+ \to p\bar{p}K^+$ mode.\cite{ppk}  
No signal is observed for $B^+ \to D^{(*)+}p\bar{p}$; the same 90\% CL upper limit $0.15 \times 10^{-4}$ is set for both modes.

\section{Charmed Hadronic Decays}
Further understanding of hadronic decay models and $s\bar{s}$ production can be gained by studying $B \to D^{*}K^-K^{(*)0}$ decays.  
BFs for these modes are presented in table~\ref{tab:dkk}.
For the $D^{(*)}K^-K^{*0}$ modes, the $K^-K^{*0}$ invariant mass and angular distributions indicate a dominant component due to $D^{(*)}a_1^+$.
$K^-K^0$ pairs in the $D^{(*)}K^-K^{0}$ modes appear to have $J^P=1^-$.

\begin{table}[t]
\caption{
  BFs and 90\% CL upper limits, in units of $10^{-4}$, for $B \to D^{*}K^-K^{(*)0}$ decays.
  \label{tab:dkk}
}
\vspace{0.2cm}
\begin{center}
\begin{tabular}{|lcccc|}
  \hline 
  Mode & $D^0 K^- K^{(*)0}$ & $D^+ K^- K^{(*)0}$ & $D^{*0} K^- K^{(*)0}$ & $D^{*+} K^- K^{(*)0}$ \\
  ${\cal B}(B \to D^{*}K^-K^{*0})$ & $7.5 \pm 1.3 \pm 1.1$ & $8.8 \pm 1.1 \pm 1.4$ & $14.0 \pm 3.1 \pm 2.6$ & $12.8 \pm 2.2 \pm 2.5$ \\
  ${\cal B}(B \to D^{*}K^-K^{0})$  & $5.5 \pm 1.4 \pm 0.8$ & $< 3.1$ & $< 11.4$ & $< 4.9$ \\
  \hline
\end{tabular}
\end{center}
\end{table}

\section{Conclusion}
An overview of the rare hadronic $B$ decay program at Belle has been presented,
including a large number of new results and first observations.
Recent results from Belle on penguin mediated $B$ decays and $B \to \eta/\omega h^{\pm}$ have been presented elsewhere.\cite{mor-ew}

%
%


\begin{thebibliography}{99}
\bibitem{kekb}    The peak luminosity of KEKB has exceeded $7 \times 10^{33} {\rm cm}^2{\rm s}^{-1}$.
\bibitem{cc}      Charge conjugate states are implied throughout.
\end{thebibliography}
\end{document}